\documentclass[preprint,5p]{elsarticle}

\usepackage{lineno,hyperref,graphicx,amsmath,amsfonts,tabularx}

\journal{Physics Letters B}

\biboptions{sort&compress}

\begin{document} \sloppy

\renewcommand{\thefootnote}{\arabic{footnote}}

\begin{frontmatter}

\title{Probing the $N = 14$ subshell closure: $g$ factor of the $^{26}$Mg(2$^+_1$) state}

\author{B.P.~M$^{\rm{c}}$Cormick}
\author{A.E.~Stuchbery\corref{cor1}}
\ead{andrew.stuchbery@anu.edu.au}
\cortext[cor1]{Corresponding author}
\author{T.~Kib{\'e}di\corref{}}
\author{G.J.~Lane}
\author{M.W.~Reed}
\author{T.K.~Eriksen}
\author{S.S.~Hota}
\author{B.Q.~Lee}
\author{N.~Palalani}
\address{Department of Nuclear Physics, RSPE, Australian National University, Canberra ACT 2601, Australia}

\begin{abstract}
The first-excited state $g$~factor of $^{26}$Mg has been measured relative to the $g$ factor of the $^{24}$Mg($2^+_1$) state using the high-velocity transient-field technique, giving $g=+0.86\pm0.10$. This new measurement is in strong disagreement with the currently adopted value, but in agreement with the $sd$-shell model using the USDB interaction. The newly measured $g$ factor, along with $E(2^+_1)$ and $B(E2)$ systematics, signal the closure of the $\nu d_{5/2}$ subshell at $N=14$. The possibility that precise $g$-factor measurements may indicate the onset of neutron $pf$~admixtures in first-excited state even-even magnesium isotopes below $^{32}$Mg is discussed and the importance of precise excited-state $g$-factor measurements on $sd$~shell nuclei with $N\neq Z$ to test shell-model wavefunctions is noted.
\end{abstract}


\end{frontmatter}


\section{Introduction}
The evolution of nuclear shell structure in exotic, radioactive neutron-rich nuclei is being studied intensively. Phenomena such as changes in shell-gap spacing and islands of inversion are revealing that nuclear structure can change significantly in neutron-rich regions \cite{OIsotopesN1416,C20shellgap,Mg32Island,ShellmodelMg32,54TiClosure,56TiN34closure,56CrClosure,54CaMagic,68NiN40Closure,usIsomers}. These changes are critical in understanding nucleon interactions and the stability of neutron-rich nuclei during nucleosynthesis \cite{54CaMagic,usIsomers,NNInteraction,pfInteraction}. The focus here is on experimental signatures of subshell closures. Usually, subshell closures are indicated first by deducing nucleon separation energies from measured masses and then, in even-even nuclei, through measurement of the energy of the first-excited $2^+$ state and its $B(E2)$ value. Of particular interest are neutron-rich nuclei near the $N=14$ \cite{OIsotopesN1416,C20shellgap}, 20 \cite{Mg32Island,ShellmodelMg32}, 34 \cite{54CaMagic} and 40 \cite{68NiN40Closure} (sub)shell closures, which exhibit unexpected shell-gap changes.

The $g$ factor of the $2^+_1$ state can be uniquely revealing of shell structure changes along an isotopic or isotonic sequence due to its dependence   on the wave-function of the specific state, and also because it is very sensitive to the occupation of single-particle orbits \cite{ShellmodelMg32,TFAnnRev,speidel02,benczerkoller07}. However, $g$-factor measurements on short-lived excited states of radioactive beams are very challenging \cite{benczerkoller07}. While experimental methods have been developed for such measurements \cite{132TeRIV,126SnTF,benczerkoller07,132TeTF,S38-40ShellStruc,3840STFRIV,136TeRIV,Zn72-Fiori}, and are yielding new insights into the structure of neutron-rich nuclei, the focus here is on the $N=14$ subshell closure in the stable nuclide $^{26}$Mg. In this case the $E(2^+_1)$ and $B(E2)$ systematics for $Z=12$ indicate a subshell closure at $N=14$: as $N$ increases from $^{22}_{12}$Mg$_{10}$ the $E(2^+)$ value spikes at $N=14$ and the $B(E2)$ value dips, indicative of a subshell closure. The expectation, then, is that the $2^+_1$ state of $^{26}$Mg should be dominated by proton excitations, giving $g(2^+_1)\sim+1$. Indeed, shell model calculations, using NuShellX \cite{NuShellX} and the USDB interactions \cite{Brown-New-USD} with the empirically optimized $M1$ operator \cite{USDA-B_Obs}, predict $g(2^+_1)=+0.959$. Surprisingly, the currently adopted value is $g(2^+_1)=+0.50(13)$ \cite{A26,SpeidMg26}, half the expected value. All experimental indicators of a shell or subshell closure should be consistent. The inconsistency of this $g$-factor measurement is therefore problematic.

The nuclide $^{26}$Mg is an example of an $sd$-shell nuclide with $N=Z+2$, the complete list being $^{18}$O, $^{22}$Ne, $^{30}$Si, $^{34}$S, and $^{38}$Ar. Within this group, the adopted experimental $g$ factors of the $2^+_1$-states in $^{18}$O, $^{22}$Ne and $^{26}$Mg are all more than two standard deviations from the theoretical values; however the case of $^{26}$Mg has the largest variance from the theoretical trend. Beyond $N=Z=12$ ($^{24}$Mg) for the magnesium isotopes, the USDB shell model must eventually break down due to intruder-state mixing \cite{USDA-B_Obs} as the island of inversion around $^{32}$Mg ($N=20$) is approached. However, a dramatic breakdown of the USDB shell model at $N=14$ is not anticipated. A new measurement of $g(2^+_1)$ in $^{26}$Mg is clearly required.

The first $g(2^+_1$,~$^{26}$Mg) measurement by Eberhardt \textit{et al.} in 1974, using the thick foil transient-field method in which the excited $^{26}$Mg ions slowed and stopped in a magnetized iron host, found $g=+0.97(18)$ \cite{EberhMg26,ZalmTF}. Later, in 1981, Speidel \textit{et al.} \cite{SpeidMg26} argued that Eberhardt \textit{et al.} had incorrectly accounted for the static-field contribution, which came into effect after the ions came to rest in the iron host. Speidel \textit{et al.} made a new measurement using the thin-foil transient-field method, which excludes the static field, and obtained $g=+0.50(13)$, in agreement with Hartree-Fock calculations available at the time. This result, which implies near equal contributions from protons and neutrons, is currently listed as the adopted value in Nuclear Data Sheets \cite{A26}. As noted above, modern shell model calculations and single-particle arguments contend that the $N=14$ subshell closure should result in $g(2^+_1)$ being much more heavily influenced by the proton contribution than the currently adopted measurement indicates. Both Eberhardt \textit{et al.} and Speidel \textit{et al.} used $(\alpha,\alpha')$ reactions to excite and recoil $^{26}$Mg ions into an iron host. The recoil velocity was relatively low, $v/c$~$\sim$~1\%, and precession angles due to the transient field were very small, $\sim$1~mrad. These were challenging experiments.

The present work reports high-velocity transient-field measurements
\cite{3840STFRIV,Zn72-Fiori} on beams of $^{24,26}$Mg ions which traversed a relatively thick ferromagnetic gadolinium host at high velocity ($v/c$~$\sim$~8\%), thus achieving precession angles that are more than an order of magnitude larger than those achieved by the $(\alpha,\alpha')$ experiments. The 2$^+_1$-state $g$~factor of $^{26}$Mg was measured relative to a recent independent and precise measurement of $g(2^+_1)$ in $^{24}$Mg \cite{Mg24RIV}.

\section{Experiment}
Transient-field $g$-factor measurements were performed using the Australian National University (ANU) Hyperfine Spectrometer \cite{ANUHIAF}. Beams of $^{24}$Mg$^{8+}$ and $^{26}$Mg$^{8+}$ at an energy of 120~MeV were produced by the 14~UD Pelletron accelerator at the ANU Heavy Ion Accelerator Facility. The beams were Coulomb excited on a cryocooled, single-layer 9.9~mg/cm$^2$ natural gadolinium target, which also served as the ferromagnetic layer for the transient-field precession effect. Calculated reaction kinematics are summarized in Table~\ref{tab:tab1}. The cryocooler kept the target at $\sim$5~K. An external magnetic field of $\sim$0.09~T was applied in the vertical direction to polarize the gadolinium foil, and was reversed every $\sim$15~min. The pole tips of the magnet were shaped to localize the polarizing field to the immediate region of the target, thus rendering the bending of the beam negligible \cite{ANUHIAF}. Calculations based on the measured field strength in the target location with the target removed show that for these Mg beams the lateral shift was $<0.5$~$\mu$m and the bending angle was $<0.3$~mrad. These values represent upper limits because the fringing field is reduced when the target foil is in place.

\begin{figure}[t]
\centering
\includegraphics[width=3.35in]{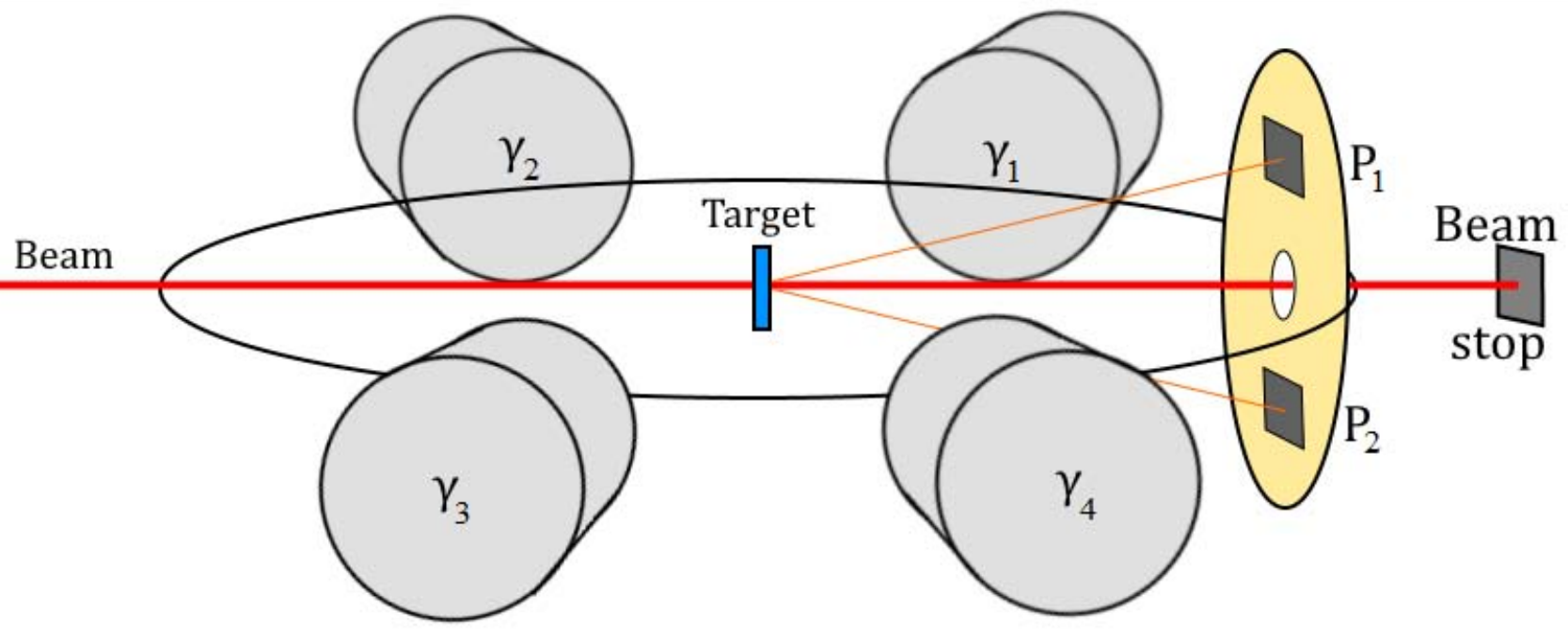}
\caption{Sketch of detector geometry (not to scale). Four NaI detectors ($\gamma_1$, $\gamma_2$, $\gamma_3$, and $\gamma_4$) were positioned around the target foil in the horizontal plane through the beam axis, while the particle detectors (P$_1$ and P$_2$) were positioned at equal angles above and below the beam axis.}\label{fig1}
\end{figure}

Four NaI detectors recorded $\gamma$~rays, and forward-scattered beam particles were detected by two 6~mm~$\times$~6~mm silicon photodiodes at an average angle of $\pm$37$^{\circ}$, centred at 18.5~mm above and below the horizontal plane. The beam intensity was kept below 2~enA, being limited by the count rate in the particle detectors. The experimental geometry is sketched in Fig.~\ref{fig1}. For the precession measurements, two $\gamma$-ray detectors ($\gamma_1$ and $\gamma_4$) were positioned in the horizontal plane at $\theta_{\gamma}=\pm 60^{\circ}$ or $\theta_{\gamma}=\pm 65^{\circ}$ while the other two ($\gamma_2$ and $\gamma_3$) were at $\theta_{\gamma}=\pm 120^{\circ}$. The angular correlation was measured for $^{24}$Mg by varying $\gamma_1$ and $\gamma_4$ through angles $\theta_{\gamma} = 0^{\circ}, \pm 15^{\circ}, \pm 30^{\circ}, \pm 45^{\circ}, \pm 55^{\circ}, \pm 60^{\circ}, \pm 65^{\circ}$, and $\pm 70^{\circ}$. For $^{26}$Mg the angular correlation was measured at $\gamma$-ray detector angles of $\theta_{\gamma}=\pm 15^{\circ}, \pm 45^{\circ}, \pm 60^{\circ}, \pm 65^{\circ}$ and $\pm 70^{\circ}$. Angular correlation data sets were normalized using a down-scaled particle count, which recorded 1~in every 1000 particle events.
\begin{table*}[t]
\centering
\caption{Average reaction kinematics for the 2$^+_1$ states of $^{24}$Mg and $^{26}$Mg traversing the gadolinium foil based on calculated Coulomb-excitation cross-sections. $E(2^+)$ is the energy of the first-excited state, $\tau(2^+)$ is the mean life of the first-excited state, $E_i$ is average energy at Coulomb excitation, $E_e$ is the average energy of exit from the foil,  $v_i$($v_e$) is the average velocity of the ion at excitation in (exit from) the foil, $\langle v\rangle$ is the average velocity of the ion in the foil, $T$ is the effective transit time, and $\Phi(\tau)$ is evaluated from Eqs. (7) and (8). $v_0 = c/137$ is the Bohr velocity. The level energies and mean lifetimes are from Refs.~\cite{A24,A26}.}
\label{tab:tab1}
\begin{tabularx}{\textwidth}{XXXXXXXXXX}
\hline
Nuclide & $E(2^+)$ & $\tau(2^+)$ & $E_i$ & $E_e$ & $v_i/v_0$ & $v_e/v_0$ & $\langle v/v_0\rangle$ & $T$ & $\Phi(\tau)$ \\
        & (keV)    &   (ps) & (MeV) & (MeV) &  &  &  & (ps) & (mrad) \\
\hline
$^{24}$Mg & 1369 & 1.92(9) & 97.0 & 61.7 & 12.8 & 10.2 & 11.5 & 0.356 & 38.7 \\
$^{26}$Mg & 1809 & 0.69(3) & 97.8 & 57.8 & 12.3 & 9.47 & 11.1 & 0.327 & 35.0 \\
\hline
\end{tabularx}
\end{table*}

The transient field induces a rotation, $\Delta\theta$, in the angular correlation, $W(\theta)$, of the $^{24,26}$Mg nuclei traversing the ferromagnetic medium, which was measured by standard procedures \cite{TFAnnRev}. Double ratios of observed counts were formed:
\begin{equation}\label{eq:rhoij}
\rho_{ij} = \sqrt{\frac{N(\theta_i)\uparrow}{N(\theta_i)\downarrow}\frac{N(\theta_j)\downarrow}{N(\theta_j)\uparrow}},
\end{equation}
where $N(\theta_i)$ and $N(\theta_j)$ represent particle-$\gamma$ coincidence counts measured in $\gamma$-ray detectors $i$ and $j$ at angles $+\theta_{\gamma}$ and $-\theta_{\gamma}$, respectively, and $\uparrow\downarrow$ represents the field direction.

The rotation angle $\Delta\theta$ is determined from:
\begin{equation}\label{eq:eps}
\epsilon = \frac{1-\rho}{1+\rho},
\end{equation}
and
\begin{equation}\label{eq:dthexp}
\Delta\theta = \frac{\epsilon}{S},
\end{equation}
where $S$ is the logarithmic derivative (``slope") of the angular correlation at $+\theta_{\gamma}$
\begin{equation}\label{eq:slope}
S = \left . \frac{1}{W}\frac{dW}{d\theta} \right |_{\theta_\gamma}.
\end{equation}

The excited $^{24}$Mg and $^{26}$Mg nuclei were allowed to recoil into vacuum after traversing the ferromagnetic layer. In this case the angular correlation of emitted $\gamma$-rays is given by \cite{ggcorrel,gpcorrel}:
\begin{equation}\label{eq:AC}
W(\theta_p,\theta_{\gamma},\Delta\phi) = \sum_{kq} B_{kq}(\theta_p) Q_k G_k F_k D^{k*}_{q0}(\Delta\phi,\theta_{\gamma},0),
\end{equation}
where $\theta_p$ and $\theta_{\gamma}$ are the particle and $\gamma$-ray detector angles (respectively), $\Delta\phi=\phi_p-\phi_{\gamma}$, $B_{kq}(\theta_p)$ is the statistical tensor defining the orientation of the nuclear state (aligned by the Coulomb excitation), $F_k$ represents the $\gamma$-ray transition $F$-coefficient \cite{Yamazaki.1967NDT3}, $D^{k*}_{q0}(\Delta\phi,\theta_{\gamma},0)$ is the rotation matrix, $Q_k$ is the finite $\gamma$-ray detector size attenuation factor, and $G_k$ is the vacuum deorientation coefficient. For our purposes, $k=0$,~2,~4. The coordinate frame is right-handed, with the beam defining the $z$-axis in the positive direction and, for our geometry, $\Delta\phi=\pi/2$ (see Fig.~\ref{fig1}). As the Mg nuclei are moving rapidly in the lab frame, the Lorentz boost must be accounted for by transforming from the lab frame to the nuclear frame \cite{gpcorrel,lorentz}.

In principle, all but the $G_k$ coefficients in Eq.~(\ref{eq:AC}) can be calculated with the required accuracy. By fitting the measured angular correlation to determine the $G_k$ values, $S$ can be determined for the evaluation of $\Delta\theta$.

The precession angle has a dependence on the level lifetime, particularly for short-lived states, which may be taken into account by expressing
\begin{equation}\label{eq:dth}
\Delta\theta = g~\Phi(\tau),
\end{equation}
where $g$ is the nuclear $g$ factor and $\Phi(\tau)$ represents the transient-field interaction for $g= 1$. $\Phi(\tau)$ is given by:
\begin{equation}\label{eq:phi}
\Phi(\tau) = -\frac{\mu_N}{\hbar} \int_{0}^{T}B_{\rm tf}[v(t)]e^{-t/\tau}dt,
\end{equation}
where $\mu_N$ is the nuclear magneton, $B_{\rm tf}[v(t)]$ is the transient-field strength at ion velocity $v(t)$, $\tau$ is the mean-life of the state of interest, and $T$ is the effective transit time of the nucleus through the ferromagnetic medium.

The transient field strength for fast ($>0.5Zv_0$), light (6 $\leq$ Z $\leq$ 16) ions traversing gadolinium hosts can be parametrized \cite{StuchTFParam} as:
\begin{equation}\label{eq:Btf}
B_{\rm tf}[v(t)] = AZ^P(v/Zv_0)^2e^{-(v/Zv_0)^4/2},
\end{equation}
where $Z$ is the atomic number of the ion and $v_0$ is the Bohr velocity. 
For gadolinium hosts, fits yield $A=26.7(11)$~T with $P=2$ fixed \cite{StuchTFParam}.

In the present measurements the same gadolinium foil serves as both target and ferromagnetic host, so the precession angle of Eq.~(\ref{eq:phi}) and all of the average kinematical quantities in Table~\ref{tab:tab1} were averaged by integrating over the energy-loss of the beam in the target and over the dimensions of the particle counters, with the integrand weighted by the Coulomb-excitation
cross section \cite{StuchMgTF,GdMag,GKINT}. The Coulomb-excitation cross section decreases by an order of magnitude as the beam loses energy through the target, so excitation occurs predominantly in the front half of the target. This method has been used previously to study high-velocity transient fields acting on Mg ions \cite{StuchMgTF}, as well as for a high-velocity transient-field $g$-factor measurement on a radioactive beam of $^{72}$Zn \cite{Zn72-Fiori}.

By combining Eq. (\ref{eq:dthexp}) and Eq. (\ref{eq:dth}), $g$-factor ratios can be determined:
\begin{equation}\label{eq:gratio}
\frac{g_x}{g_y} = \frac{\epsilon_x}{\epsilon_y} \frac{S_y}{S_x} \frac{\Phi_y}{\Phi_x} = \frac{\Delta\theta_x}{\Delta\theta_y} \frac{\Phi_y}{\Phi_x},
\end{equation}
where $x$ and $y$ signify the two states being measured.

\begin{figure}
\centering
\includegraphics[width=3.35in]{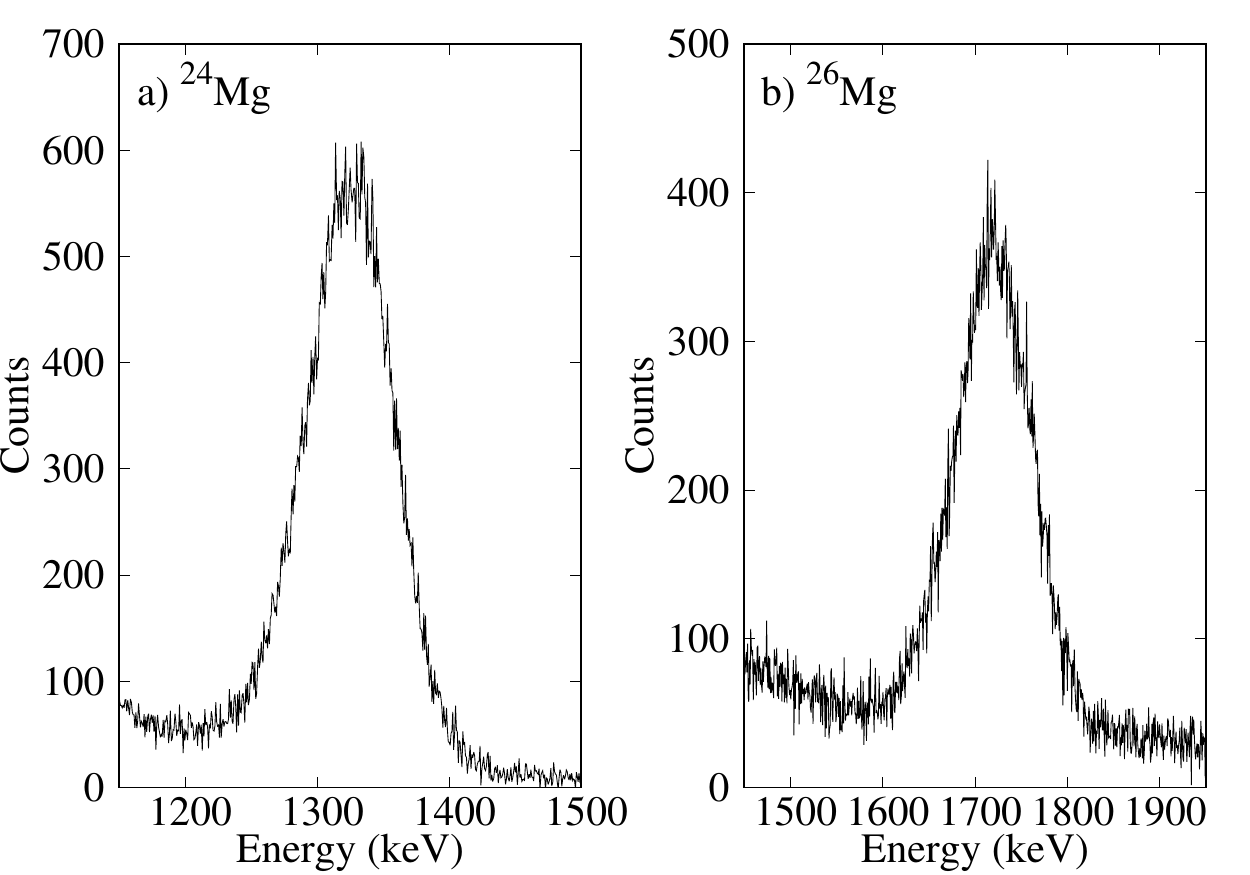}
\label{fig2}
\caption{Photopeak region of the random-subtracted particle-$\gamma$ coincidence spectra observed in $\gamma_2$ ($120^{\circ}$) for a) $^{24}$Mg and b) $^{26}$Mg. The spectra show the field-up, field-down, P$_1$, and P$_2$ data summed across all runs.}
\end{figure}

\section{Results and Analysis}

Examples of random-subtracted $\gamma$-ray spectra in coincidence with particles are shown in Fig.~\ref{fig2}. A particle-gamma coincidence $\gamma$-ray spectrum taken with a HPGe detector indicated that the regions of interest (1369 keV and 1809 keV) had no contamination after random subtraction.

\begin{figure}
\centering
\includegraphics[width=3.35in]{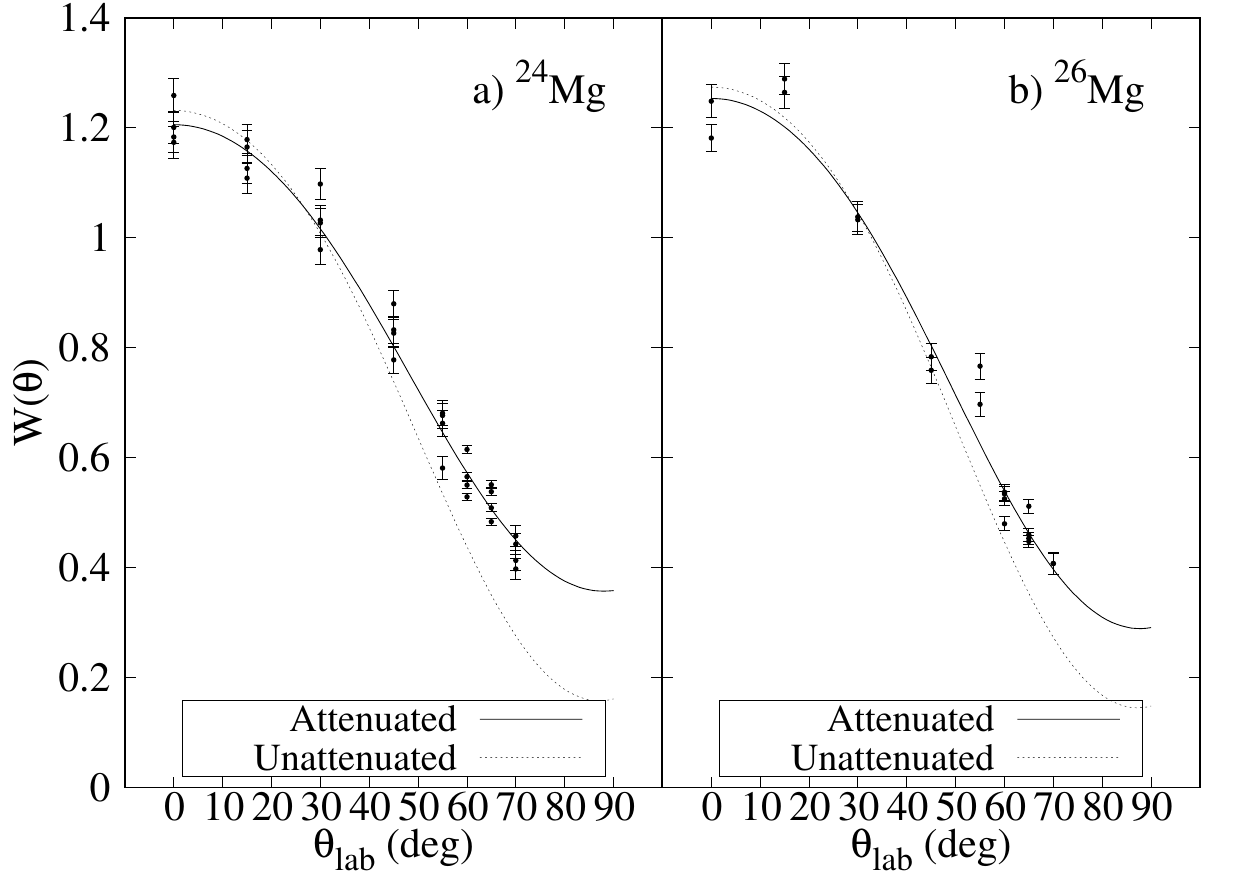}
\label{fig3}
\caption{Angular correlations in the laboratory frame for a) $^{24}$Mg and b) $^{26}$Mg. The data are shown along with the calculated unattenuated correlation (dotted line) and the fit that is attenuated by vacuum deorientation (solid line).}
\end{figure}

The lab-frame angular correlation data shown in Fig.~\ref{fig3} were fitted to determine $G_2$ and $G_4$, and hence deduce $S$ values. As the $G_2$ and $G_4$ parameters are highly correlated for the available data, they were related through a single $J$ = 1/2 electron-spin (H-like) fraction parameter, as described in a previous study of high-velocity $^{24}$Mg ions \cite{StuchMgTF}, which used a methodology similar to that of the present measurement. Fits returned a $J$ = 1/2 fraction of $\sim$50\%, which agrees with calculations of charge-state distributions using the Schiwietz-Grande formula \cite{SchiwietzGrande}, summing the H-like and Li-like contributions. The $S$ values so obtained agree well with those obtained allowing $G_2$ and $G_4$ to vary freely, but avoided the complications of handling the errors on correlated parameters. While the $S$ values for the forward-placed detectors at $\theta_{\gamma} = \pm60^{\circ}$ and $\pm65^{\circ}$ could be determined from the fit to measured angular correlations, those for the backward detectors at $\theta_{\gamma} = \pm 120^{\circ}$ were inferred from the fit to the measured angular correlations at forward angles, the difference between $\pm60^{\circ}$ and $\pm120^{\circ}$ originating only from the effect of the Lorentz boost.

Measured precession angles are listed in Table~\ref{tab:tab2}. The relative $g$~factors were determined from Eq. (\ref{eq:gratio}) as
\begin{equation}
\nonumber
  \frac{g(2^+_1; ^{26}\rm{Mg})}{g(2^+_1; ^{24}\rm{Mg})} = \frac{26.9(21)}{18.6(16)} \times \frac{38.7}{35.0} =  1.60(19).
\end{equation}
Taking $g(2^+_1; ^{24}\rm{Mg}) = +0.538(13)$ gives $g(2^+_1; ^{26}\rm{Mg}) = +0.86(10)$.
Note that a 2.4\% uncertainty (with no significant impact on the uncertainty in the $g$ factor) was assigned to the ratio $\Phi(^{24}{\rm Mg})/\Phi(^{26}{\rm Mg}) =38.7/35.0$ to account for uncertainty in the velocity-dependence of the transient field. This uncertainty was estimated by comparing this adopted ratio based on  Eq. (\ref{eq:Btf}) to an evaluation of $\Phi(^{24}{\rm Mg})/\Phi(^{26}{\rm Mg})$ under the assumption that $B_{\rm tf} \propto v$.
The $g$-factor measurement is effectively independent of the assumed velocity dependence of the transient field because both level lifetimes are longer than the transit time through the gadolinium foil (see Table \ref{tab:tab1}).

The experimental value of $\Phi_{\rm exp}(^{24}{\rm Mg})=\Delta\theta / g = 35(3)$~mrad, is in agreement with the parametrization of Eq.~(\ref{eq:Btf}) (see Table~\ref{tab:tab1}), considering that uncertainties in the gadolinium target thickness ($\sim5\%$) have been ignored, and that a reduced magnetization is often found for such relatively thick gadolinium foils \cite{GdMag}.

Precession angles an order of magnitude larger than the earlier works \cite{EberhMg26,SpeidMg26} were observed in the present measurement. Moreover, the same target was used with beam excitation to measure the ratio of $2^+_1$-state $g$ factors in $^{24}$Mg and $^{26}$Mg. As such, the $g$-factor ratio is determined essentially by the ratio of the `effects' $\epsilon$, with relatively small corrections due to differences in $S$ (arising from differences in vacuum deorientation), and effective transient-field strengths, which largely cancel [see Eq.~(\ref{eq:gratio}) and Table~\ref{tab:tab2}]. These features of the experiment help ensure a robust and reliable result.

\begin{table}[]
\centering
\caption{Experimental results}
\label{tab:tab2}
\begin{minipage}[t][3.7cm]{\hsize}
\renewcommand{\thefootnote}{\alph{footnote}}
\begin{tabularx}{\hsize}{ccccc}
\hline
Nuclide & $\pm\theta_{\gamma}$ & $\epsilon\times$10$^3$ & $S$ [rad$^{-1}$] & $\Delta\theta$ (mrad) \\
\hline
$^{24}$Mg & 60 & +23.3(35) & $-$1.299(26) & $-$18.0(27) \\
 & 65 & +22.8(69) & $-$1.294(26) & $-$17.6(54) \\
 & 120 & $-$23.5(25) & +1.229(25) & $-$19.2(21) \\
 & & & & \textbf{$-$18.6(16)}$^a$ \\
$^{26}$Mg & 60 & +35.5(126) & $-$1.573(31) & $-$22.6(80) \\
 & 65 & +46.9(53) & $-$1.618(32) & $-$29.0(33) \\
 & 120 & $-$37.7(39) & +1.455(30) & $-$25.9(28) \\
 & & & & \textbf{$-$26.9(21)}$^a$ \\
\hline
\end{tabularx}
$^a$Weighted average.
\end{minipage}
\end{table}

The present $g$-factor measurement agrees with that of Eberhardt \textit{et al.} \cite{EberhMg26}, but with a reduced uncertainty. It appears that the transient-field calibration and the magnitude of the static-field contribution, which were questioned by Speidel \textit{et al.} \cite{SpeidMg26}, were appropriately handled by Eberhardt \textit{et al.} after all. Our  result disagrees with that of Speidel \textit{et al.}, who reported similar transient-field precession angles for both $^{24}$Mg and $^{26}$Mg. A careful examination of their publication did not indicate any particular reason for the disagreement with our work, although it is possible that their $^{26}$Mg target had a thinner iron layer than reported. We offer this suggestion because the measurements on $^{24}$Mg in iron reported by Speidel \textit{et al.} \cite{SpeidMg26} seem to agree with other independent measurements, and correspond to expected $B_{\rm tf}$ values for Ne, Mg and Si ions traversing iron at similar ion velocities \cite{ZalmTF,EberhardtFeFoil}.

\section{Discussion}

\begin{figure}
\centering
\includegraphics[width=3.35in]{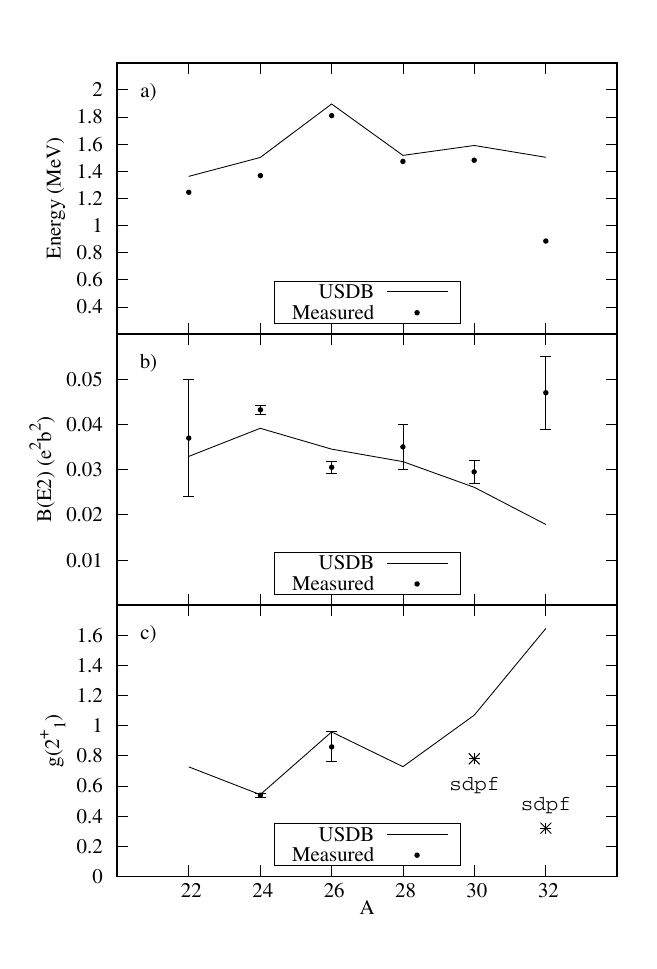}
\label{fig4}
\caption{Comparison of USDB shell model calculations and experiment for the magnesium isotopes from $A=22$ to 32 a) $E(2^+_1)$ energies, b) $B(E2)$ rates, and c) $g$-factor values \cite{RamanBE2,A22,A24,A26,A28,A30,A32}. The theoretical $g$ factors for $^{30}$Mg and $^{32}$Mg in a more realistic $sdpf$ model space are also shown by the stars \cite{MCShellModel}.}
\end{figure}

The $E(2^+_1)$, $B(E2)$ and $g(2^+_1)$ systematics of the even-$A$ magnesium isotopes from $^{22}$Mg to $^{32}$Mg are shown in Fig.~\ref{fig4}. These values show a spike in the $E(2^+_1)$ value and a dip in the $B(E2)$ value at $^{26}$Mg. Together, these two features are indicative of a subshell closure in $^{26}$Mg. Specifically, the $\nu d_{5/2}$ subshell is filled. Shell-model calculations performed with NuShellX \cite{NuShellX} and the USDB interaction \cite{Brown-New-USD,USDA-B_Obs} indicate the $g(2^+_1)$ of $^{26}$Mg to be almost double that of neighbouring $^{24}$Mg, and in agreement with our measured value at the level of one standard deviation. The calculated spin decompositions of the $2^+_1$ states in $^{24}$Mg and $^{26}$Mg, listed in Table~\ref{tab:tab3}, show a strong single-proton influence in the $^{26}$Mg($2^+_1$) state. The behaviour of the leading terms indicates the behaviour of the $g$ factors: For $^{24}$Mg the $2^+_1$ state has equal (26\%) components of $\nu(2^+)\otimes\pi(0^+)$ and $\nu(0^+)\otimes\pi(2^+)$, whereas in $^{26}$Mg the $\nu(0^+)\otimes\pi(2^+)$ component is dominant (52\%) and $\nu(2^+)\otimes\pi(0^+)$ is much smaller (17\%).

Although the present results are in agreement with the USDB shell model, the model must break down as $^{32}$Mg and the so-called island of inversion is approached \cite{ShellmodelMg32,MCShellModel}. As indicated in Fig.~\ref{fig4}, for $^{32}$Mg the USDB interaction in the $sd$ model space gives $g(2^+_1)=+1.6$ whereas more realistic Monte Carlo Shell Model calculations in a $sdpf$ model space by Otsuka {\em et al.} \cite{MCShellModel} give $g=+0.32$, very much smaller than the $sd$-shell model value. In $^{30}$Mg the $g(2^+_1)$ value in the $sdpf$ space remains $\sim 20\%$ smaller than the $sd$-model value \cite{MCShellModel}.

\begin{table}[]
\centering
\caption{Spin composition of $2^+_1$ states in $^{24,26}$Mg}\label{tab:tab3}
\begin{tabularx}{\columnwidth}{XXXX}
\hline
$J_n$ & $J_p$ & \multicolumn{2}{c}{Weight (\%)} \\
\cline{3-4} \\
 & & $^{24}$Mg & $^{26}$Mg \\
\hline
2 & 0 & 25.64 & 17.05 \\
0 & 2 & 25.64 & 52.04 \\
2 & 2 & 19.66 & 9.59 \\
2 & 4 & 8.60 & 7.17 \\
4 & 2 & 8.60 & 3.85 \\
\hline
\end{tabularx}
\end{table}

The precisely measured ground-state $g$ factors of the odd-$A$ magnesium isotopes, at face value, might suggest a rather abrupt transition to the island of inversion \cite{USDA-B_Obs,Mg31meas}. However, the ground-state moments are relatively insensitive to configuration mixing across the $N=20$ shell gap because they are largely determined by the odd neutron, not the behaviour of the core. The case of $^{31}$Mg illustrates this point: The measured $1/2^+$ ground-state moment \cite{Mg31meas} is reasonably well described by USDB shell-model calculations, but the predicted $1/2^+$ state is at an excitation energy above 2~MeV and is not the ground state \cite{USDA-B_Obs}.

Studies of the excited-state spectroscopy of $^{30}$Mg have shown that the $sd$-shell model fails at moderate spin, and cross-shell ($pf$) excitations are needed at rather low excitation energy \cite{Mg30Ex}. Certainly, the $2^+_1$ states must be expected to contain more $pf$ admixtures than the ground states, and $g(2^+_1)$ values may show a smoother transition to the island of inversion than the ground-state moments of the odd-$A$ isotopes. Thus, although the experimental uncertainty is too large to draw conclusions, the fact that the present $g(2^+_1)$ result for $^{26}$Mg tends to fall below the USDB prediction is intriguing. It invites a more precise $g$-factor measurement on the $^{26}$Mg $2^+_1$ state, and also on neutron-rich $^{28}$Mg $2^+_1$, which could be achieved by use of the time-dependent recoil in vacuum (TDRIV) method, as applied recently to $^{24}$Mg $2^+_1$ \cite{Mg24RIV}. (Although the RIV method gives only the magnitude of the $g$~factor, it has proven to give it more precisely than the transient-field method \cite{Mg24RIV}, particularly in the case of radioactive beam measurements where statistical precision is limited; compare Refs.~\cite{Allmond-Sn126RIV,Sn126TF}. The primary reason is that the transient-field method requires $\gamma$-ray detection at a few specific angles in the plane perpendicular to the direction of the applied magnetic field whereas the RIV method can take advantage of $\gamma$-ray detection over a much broader angular range. A second reason, applicable for hydrogen-like Mg ions \cite{Mg24RIV}, is that the hyperfine interaction of the free ion in vacuum  can be calculated from first principles with very high accuracy.)

\begin{figure}
\centering
\includegraphics[width=3.35in]{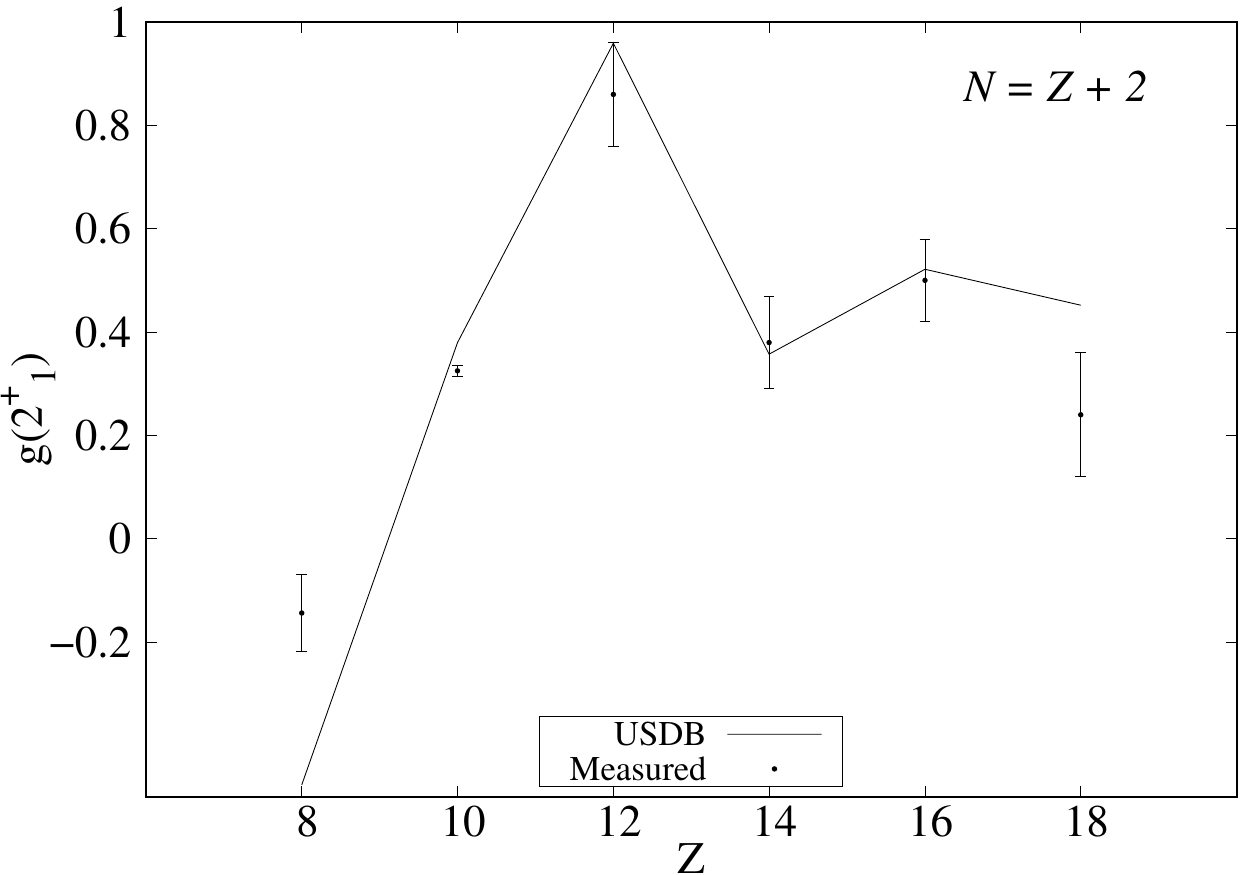}
\caption{Measured and USDB shell-model calculated $g$ factors for $N=Z+2$ $sd$-shell nuclei \cite{A18,A22,A26,A30,A34,A38}.}
\label{fig5}
\end{figure}

Finally, returning to the $g$ factors of the $sd$-shell nuclei with $N = Z + 2$, which are displayed in Fig.~\ref{fig5}, it is evident that with the new result for $g(2^+_1)$ in $^{26}$Mg, the experimental and theoretical trends are in agreement. The experimental values for $^{18}$O and $^{22}$Ne, however, remain over two standard deviations from theory. Further investigation is needed to determine whether these discrepancies are due to the experimental data, or signal a short coming in the USDB shell-model wavefunctions.

In summary, the $g$ factor of the first-excited state in $^{26}$Mg has been measured by the high-velocity transient-field method. Conflicting previous values from very low-velocity transient-field measurements \cite{EberhMg26,SpeidMg26} are perhaps best set aside, however the new measurement agrees with the measurement of Eberhardt \textit{et al.} \cite{EberhMg26,ZalmTF}. It also agrees with USDB shell-model calculations, but does not exclude the possibility that $g(2^+_1)$ in $^{26}$Mg may begin to reduce from the USDB model due to emerging neutron $pf$ admixtures, which must become prominent as the magnesium isotopes approach $N=20$ \cite{Mg32Island,ShellmodelMg32,MCShellModel,Mg30Ex}. In any case, the excited-state $g$~factors of $sd$-shell nuclei with $N=Z+2$ are more sensitive to the proton-neutron balance in the wavefunctions than in nuclei with $N=Z$, where $g\simeq$~0.5 in all cases. Efforts to improve the precision and accuracy of experimental $g(2^+_1)$ values in nuclei with $N\neq Z$ can therefore provide new opportunities to test the wavefunctions of the $sd$-shell model.

\section*{Acknowledgements}
The authors are grateful to the academic and technical staff of the Department of Nuclear Physics and the Heavy Ion Accelerator Facility (Australian National University) for their continued assistance and maintenance of the facility. We thank Dr. G. Georgiev for thoughtful comments on the manuscript.
This research was supported in part by the Australian Research Council grant numbers DP120101417, DP130104176, DP140102986, DP140103317 and FT100100991. B.P.M. acknowledges the support of the Australian Government Research Training Program. Support for the ANU Heavy Ion Accelerator Facility operations through the Australian National Collaborative Research Infrastructure Strategy (NCRIS) program is acknowledged.



\end{document}